\documentclass[preprint2]{aastex}

\shorttitle{Low Order WFS for Coronagraphic Systems}
\shortauthors{O. Guyon, T. Matsuo, R. Angel}
\usepackage{graphicx}
\usepackage[mathscr]{eucal}
\begin{document}

\title{Coronagraphic Low Order Wavefront Sensor: Principle and Application to a Phase-Induced Amplitude Coronagraph}
  
\author{Olivier Guyon}
\affil{National Astronomical Observatory of Japan, Subaru Telescope, Hilo, HI 96720}

\author{Taro Matsuo}
\affil{NASA Jet Propulsion Laboratory, Pasadena , CA 91109}

\author{Roger Angel}
\affil{Steward Observatory, University of Arizona, Tucson, AZ 85719}

\email{guyon@naoj.org}

\begin{abstract}
High contrast coronagraphic imaging of the immediate surrounding of stars requires exquisite control of low-order wavefront aberrations, such as tip-tilt (pointing) and focus. 
We propose an accurate, efficient and easy to implement technique to measure such aberrations in coronagraphs which use a focal plane mask to block starlight. 
The Coronagraphic Low Order Wavefront Sensor (CLOWFS) produces a defocused image of a reflective focal plane ring to measure low order aberrations. 
Even for small levels of wavefront aberration, the proposed scheme produces large intensity signals which can be easily measured, and therefore does not require highly accurate calibration
of either the detector or optical elements. The CLOWFS achieves nearly optimal sensitivity and is immune from non-common path errors.
This technique is especially well suited for high performance low inner working angle (IWA) coronagraphs. On phase-induced amplitude apodization (PIAA) type coronagraphs, it can unambiguously recover aberrations which originate from either side of the beam shaping introduced by the PIAA optics.
We show that the proposed CLOWFS can measure sub-milliarcsecond telescope pointing errors several orders of magnitude faster than would be possible in the coronagraphic science focal plane alone, and can also accurately calibrate residual coronagraphic leaks due to residual low order aberrations.
We have demonstrated $\approx 10^{-3} \lambda/D$ pointing stability in a laboratory demonstration of the CLOWFS on a PIAA type coronagraph.
\end{abstract}
\keywords{instrumentation: adaptive optics --- techniques: high angular resolution}

\section{Introduction}

High performance coronagraphs with small inner working angle (IWA) are unavoidably very sensitive to small pointing errors and other low order aberrations \citep{lloy05,shak05,siva05,beli06,guyo06}. This property is due to the fact that the wavefront of a source at small angular distance (typically between 1 and 2 $\lambda/D$) from the optical axis is "similar" (in the linear algebra sense of the term) to an on-axis wavefront with a small ($\ll \lambda/D$) pointing error. If the coronagraph must "transmit" the former, it will also transmit a significant part of the latter, and therefore be extremely sensitive to pointing errors. This behavior is indeed verified by performance comparison between coronagraph concepts \citep{guyo06}.

\begin{figure*}[htb]
\includegraphics[scale=0.35]{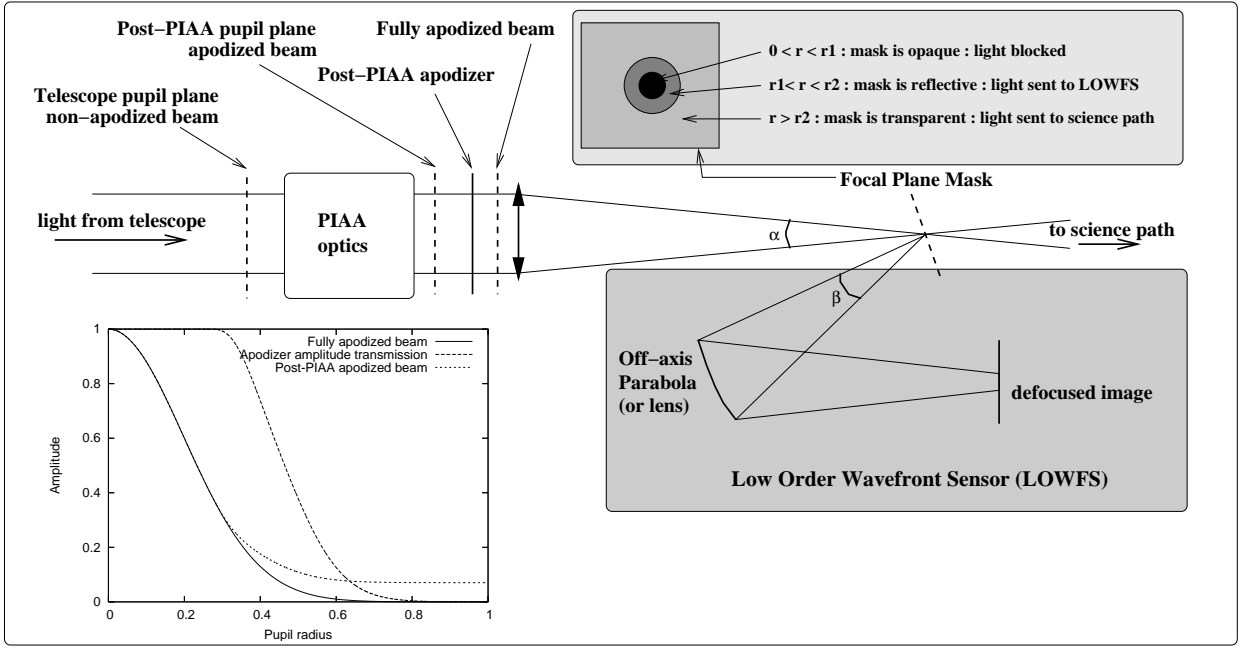} 
\caption{\label{fig:lowfsprinciple} Optical layout of a coronagraphic low order wavefront sensor system, shown here with a PIAA coronagraph. See text for details.}
\end{figure*}

For high performance coronagraphs, such as the Phase-Induced Amplitude Apodization (PIAA) coronagraph \citep{guyo03,guyo05} used as example in this paper, the coronagraph should ideally be designed to balance IWA against stellar angular diameter, which sets a fundamental limit on the achievable coronagraph performance. Such coronagraphs are therefore pushed to become sensitive to pointing errors corresponding to the angular size of nearby stars, roughly 1 milliarcsecond (mas), and are also highly sensitive to other low order wavefront errors such as focus and astigmatism. Milliarcsecond-level pointing error can increase stellar leakage in the coronagraph to the point where a planet would be lost in photon noise. Even smaller errors can create, if not independantly measured, a signal which is very similar to a planet's image.

A robust and accurate measurement of low order aberrations (especially tip-tilt errors, which are easily generated by telescope pointing errors and vibrations) is therefore essential for high contrast coronagraphic observations at small angular separation. The science focal plane after the coronagraph is unfortunately "blind" to small levels of low order aberrations, which can only be seen when already too large to maintain high contrast in the coronagraphic science image. A better option is to monitor pointing errors by using starlight which would otherwise be rejected by the coronagraph. This scheme was successfully implemented on the LYOT project coronagraph \citep{oppe04,digb06}. Alternatively, the measurement could be performed independently from the coronagraph optical train (for example, the wavefront sensor in the adaptive optics system upstream of the coronagraph).
We propose in this paper an improved solution to obtain accurate measurement of several low-order aberrations including pointing: the coronagraphic low-order wavefront sensor (CLOWFS). The wavefront control requirements for a PIAA coronagraph are first clearly defined in \S\ref{sec:requ}. The CLOWFS principle is presented in \S\ref{sec:principle}. Wavefront reconstruction algorithms and CLOWFS sensitivity are discussed in \S\ref{sec:wfreconstr}. The aberration sensitivity of a PIAA coronagraph equipped with a CLOWFS is discussed in \S\ref{sec:aberrsens}, and the results of a laboratory demonstration on a PIAA Coronagraph system are shown in \S\ref{sec:labexp}.

\section{Low order wavefront control requirements for PIAA coronagraphs}
\label{sec:requ}
\begin{figure}[htb]
\includegraphics[scale=0.8]{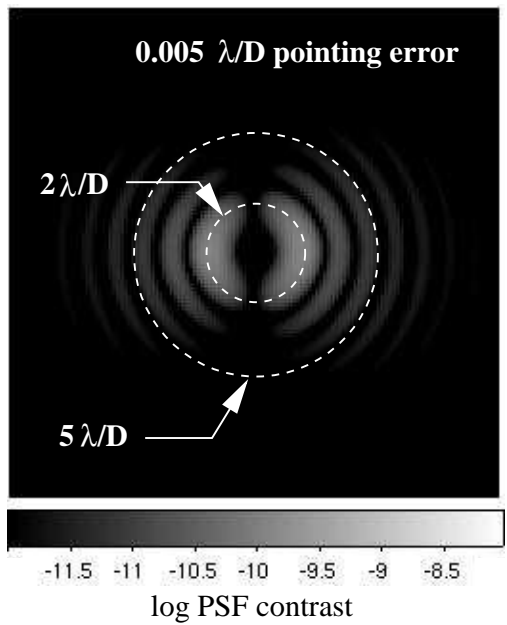}
\caption{\label{fig:psf21} Simulated science focal plane image in a PIAAC system with a 0.005 $\lambda$/D pointing offset. The PIAA design adopted for this simulation has a 1.9 $\lambda$/D IWA.}
\end{figure}
Wavefront aberrations can be produced either before the PIAA optics (for example a bending of the telescope primary mirror, or a telescope pointing error) or between the PIAA optics and the focal plane mask.
In a space-based telescope free of atmospheric turbulence, the strongest sources of aberration are likely to be telescope pointing errors and low-order aberrations due to structural deformations of the optical telescope assembly. The PIAA coronagraph is especially sensitive to such low order aberrations if they are introduced prior to the PIAA optics \citep{beli06}, as they will then scatter light in the most scientifically precious area of the science focal plane: the inner part of the field. As illustrated in Figure \ref{fig:psf21}, a very small pointing error (0.5\% of $\lambda$/D) can be sufficient to create an artefact as bright as an Earth-like planet. Extreme sensitivity to pointing is unavoidable in small IWA coronagraphs, and, in the case of the PIAA, is due to the fact that a "remapped" tip/tilt scatters light outside the focal plane mask. Coronagraphs with larger IWA and better tolerance to pointing errors exist, and even within the PIAA "family" of coronagraph, sensitivity to pointing errors can be balanced against IWA by changing the size of the focal plane mask and the pupil apodization profile. 

The PIAA coronagraph is much less affected by aberrations after the beam shaping optics. For example, a small amount of tip/tilt in the post-PIAA optics will simply move the PSF on the focal plane mask (which is most likely slightly oversized) without introducing scattering at larger separations. In addition, low order aberrations introduced after the PIAA optics will likely be much smaller in amplitude, thanks to the small size of optical elements between the PIAA optics and the focal plane mask.

For successful detection of a faint source (assumed here to be an Earth-like planet at $10^{-10}$ contrast), low-order aberrations must simultaneously:
\begin{itemize}
\item{Be small enough to avoid loosing the planet signal within the scattered starlight's photon noise. For the PIAA coronagraph design used in this work, a pre-PIAA 0.005 $\lambda$/D pointing error is sufficient to scatter $6.9 \times10^{-10}$ of the starlight into the science field, and this scattered light produces two wide "arcs" on either side of the optical axis, at a $10^{-10}$ contrast level. Although a PIAA coronagraph could be designed with reduced sensitivity to pointing errors (but with larger IWA), we assume here that the maximum allowable pointing error for a space PIAA coronagraph mission aimed at direct imaging of "Earth-like" planets is 0.005 $\lambda$/D (corresponding to starlight leak peaking at $10^{-10}$ contrast). On a 1.4-m diameter telescope in the visible, this corresponds to 0.4 mas, a value which is similar to the angular radius of a "typical" target (a main-sequence star at 10 pc). On larger telescopes, the angular radius of the star (0.5 mas for a Sun-like star at 10 pc) will drive the coronagraph design, which will therefore also end up with a $\approx$ 0.5 mas RMS pointing error requirement. }
\item{Be stable, or calibrated, to a fraction of the planet's expected flux. We assume in this paper that coronagraphic leaks due to low order aberrations must be calibrated to 10\% of the expected planet's contribution. This second requirement is therefore much more severe. Light scattered in the science focal plane scales approximately as the square of the aberration amplitude: pointing must be stable (or calibrated) to $\approx 0.0016 \lambda/D$ (0.13 mas on a 1.4-m telescope in the visible) for a planet at $10^{-10}$ contrast. Fortunately, high accuracy measurement of low order aberrations by the CLOWFS scheme proposed in this paper can be used to reliably model the stellar leakage in the coronagraphic science focal plane. This model can then be numerically subtracted from the science image to reveal much fainter underlying sources down to the photon noise and detector noise limits. Pointing errors larger than 0.0016 $\lambda/D$ (but smaller than 0.005 $\lambda/D$) are acceptable, as long as they are measured to 0.0016 $\lambda/D$ accuracy.}
\end{itemize}

\section{Coronagraphic Low Order Wavefront Sensor: Principle}
\label{sec:principle}

\subsection{Optical Layout}
A simplified optical layout for a CLOWFS system for a PIAA coronagraph is shown in Figure \ref{fig:lowfsprinciple}. CLOWFS light is extracted by the focal plane mask located after the PIAA optics. In the PIAA coronagraph design, the role of this focal plane is to selectively remove starlight, while transmitting the science field. The mask is therefore illuminated by a large number of photons, which are freely available, and, if properly used, allow highly sensitive measurement of low order aberrations.

\begin{figure}[htb]
\includegraphics[scale=0.6]{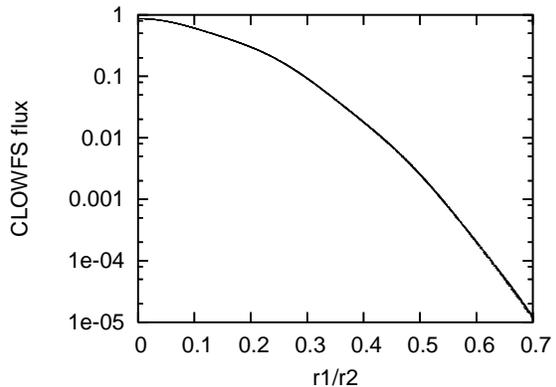} 
\caption{\label{fig:fluxr1r2} Fraction of the total starlight collected by the telescope sent to the CLOWFS detector as a function of $r_1/r_2$.}
\end{figure}

The central part of the focal plane mask used in the CLOWFS design is opaque: only a reflective annulus around this central part sends light to the CLOWFS optics. As shown in figure \ref{fig:lowfsprinciple} we denote $r_1$ the radius of the inner opaque zone and $r_2$ the radius of the focal plane mask, which is fixed at the IWA of the coronagraph. For the baseline configuration adopted in this paper $r_1/r_2 = 0.4$, with $r_2 \approx 1.8 \lambda/D$ on the sky: the central 40\% of the focal plane mask is opaque. Since the pupil before the focal plane mask is strongly apodized by the PIAA optics, only a small fraction of the total light is reflected by the reflective annulus covering the $r_1<r<r_2$ range (see Figure \ref{fig:fluxr1r2}). 

The CLOWFS detector acquires a defocused image of this annulus, at defocus distance $20 rad = 3.2$ waves (in this paper, the defocus value is given as Peak-to-Valley in the post-PIAA pupil plane). While this defocus value may seem large, most of the light is in the central part of the apodized pupil, and the "effective" defocus is about half to a third of the optical defocus listed in this work. The CLOWFS pupil is oversized by a factor 2 ($\beta = 2 \alpha$) to include most of the light reflected by the CLOWFS focal plane mask.

\subsection{Importance of defocusing the CLOWFS}
The CLOWFS defocus (the CLOWFS detector is not conjugated to the focal plane mask) is necessary to measure focus, but is not necessary if only tip and tilt are measured.
In a CLOWFS configuration where the detector is conjugated to the focal plane mask, a single CLOWFS image can be used to recover the amplitude of focus in the incoming beam, but not its sign (in an optical system free of aberrations, focal plane images acquired inside and outside focus are identical). The CLOWFS defocus is therefore introduced to remove this sign ambiguity. A mathematical proof of the quadratic (rather than linear) response of a "in-focus" CLOWFS to pupil focus aberration is provided in \S\ref{ssec:lin}.

\subsection{Why a dual zone focal plane mask ?}
While in an ideal system (where photon noise would be the only source of noise) the central opaque zone would reduce the CLOWFS performance, it is in practice extremely advantageous for two reasons, which are now described.
\subsubsection{Relative signal amplification}
\begin{figure}[htb]
\includegraphics[scale=0.45]{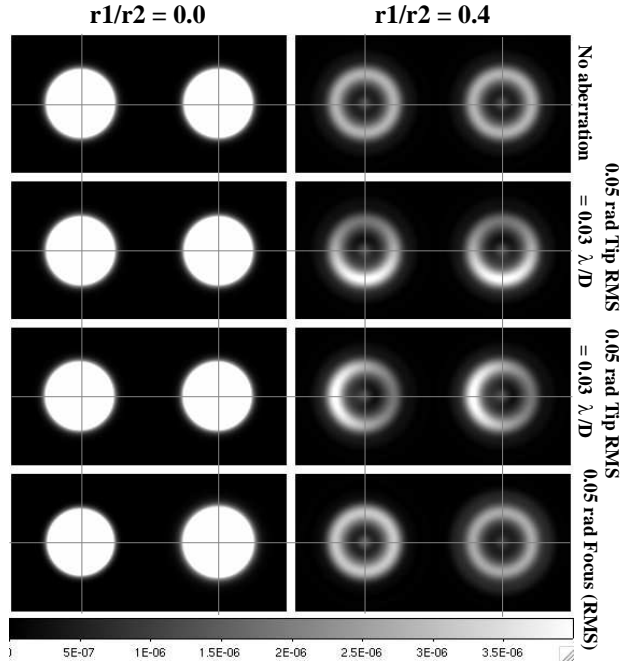} 
\caption{\label{fig:lowfsims} Images obtained by the CLOWFS for $r_1/r_2 = 0$ (left) and $r_1/r_2 = 0.4$ (right) for a defocus distance of $20 rad$. With $r_1/r_2 = 0.4$, the images are significantly fainter but aberrations are more easily seen. In each of the 8 images shown, the inside and outside focus images of the CLOWFS mask are side-by-side. In the CLOWFS system we propose, only one of these two images would need to be acquired.}
\end{figure}

Without the opaque zone, the CLOWFS detector would be illuminated by a large number of photons and the signal to be extracted would be a tiny relative change in intensity, which would be challenging to measure in practice. To detect this signal, the detector calibration would need to be very accurate and detector saturation would be a serious issue. Masking the central part of the PSF greatly reduces the total amount of light in the CLOWFS but only slightly reduces the amplitude of the signal produced by low order aberrations. For example, a small pointing error in the post-PIAA pupil creates maximal signal where the PSF surface brightness slope is the greatest. The very center of the PSF, although it contains a lot of flux, contains very little signal. Thanks to the central dark zone, microscopic changes in the coronagraphic wavefront produce macroscopic changes in the CLOWFS image, as shown in Figure \ref{fig:lowfsims}.

\subsubsection{Self-referencing}
The central dark zone of the focal plane mask allows the CLOWFS to be insensitive to small motions of the CLOWFS elements (detector and off-axis parabola or lens to re-image the focal plane mask).

If the focal plane mask were fully reflective, a tip-tilt error in the post-PIAA pupil (which should be corrected) and a tip-tilt error in the CLOWFS optics (which should not be corrected) would look identical. This is due to the fact that the outer edge of the focal plane mask is dark, and the CLOWFS would therefore measure tip-tilt as a translation of the image on the CLOWFS detector. This sensitivity to small amount of tip-tilt in the CLOWFS optics would therefore require the CLOWFS optical elements positions to be accurately known and stable. With the dark zone in the center of the focal plane mask, however, the same tip-tilt is measured as a macroscopic change in the CLOWFS image shape (see Figure \ref{fig:lowfsims}) instead of a small translation. Thanks to the dark zone, the CLOWFS tip-tilt measurement is accurately referenced to the focal plane mask.

\subsection{Sensitivity to focal plane mask manufacturing errors}

The geometry of the focal plane mask is driven by both the coronagraph architecture and the CLOWFS. 
The transmission of the mask must first satisfy the coronagraph's requirements. Implementing a CLOWFS requires:
\begin{itemize}
\item{A tilted focal plane mask to redirect stellar light to the CLOWFS imaging optics. If the tilt angle is large, or if the coronagraph requires a very circular focal plane mask, the focal plane mask could be made elliptical.}
\item{A reflective zone on the focal plane mask. This requirement should have no negative impact on the coronagraph's performance, as a reflective coating may be deposited on the illuminated side of the focal plane mask without affecting its transmission.}
\end{itemize}

The focal plane mask reflectivity defines the CLOWFS performance. An "ideal" focal plane mask for CLOWFS, as described in this paper, may be very challenging to manufacture:
\begin{itemize}
\item{It is very difficult to make the central part of the mask truly "black", and some of the light in the central part of the PSF will be reflected into the CLOWFS imaging optics.}
\item{The reflective annulus may not be uniformly reflective and could also introduce wavefront errors (the reflective surface may not be very flat).}
\item{Coatings, wether optimized to be black (central part of the mask) or reflective (annulus), are somewhat chromatic: the reflectivity map of the focal plane mask is wavelength-dependent.}
\end{itemize}
Even with these imperfections, the CLOWFS will still produce a strong response for the aberrations to be measured (tip, tilt, focus and their remapped equivalents). It may however be difficult to predict what there responses will be if the focal plane mask is poorly calibrated. As detailed in \S\ref{sec:wfreconstr}, we therefore propose to first measure these linear responses by introducing aberrations in the coronagraph optics, and then use these responses to decompose the CLOWS image in a linear sum of aberrations. This step requires no additional hardware, provided that there are actuators to correct for the aberrations measured by the CLOWFS. This is very similar to what is commonly done on adaptive optics systems, where a "response matrix" is first acquired by measuring the WFS signal when each actuator of the deformable mirror is moved.

\section{Wavefront reconstruction algorithm and sensitivity}
\label{sec:wfreconstr}

\subsection{Linearity}
\label{ssec:lin}
We show in this section that for small wavefront aberrations ($<<$ 1 rad), the CLOWFS image is a linear function of the wavefront aberration modes to be measured. This convenient property is due to:
\begin{itemize}
\item{The fact that the CLOWFS is not operating on a "dark" fringe: when the incoming wavefront is perfect, the CLOWFS images already contain some light.}
\item{The non-orthogonality between the aberration-free CLOWFS complex amplitude distribution in the CLOWFS detector array and the change introduced on this complex amplitude by the aberration modes to be measured.}
\end{itemize}
We denote $u$ and $x$ the 2-D coordinate respectively in the post-PIAA pupil plane and in the CLOWFS detector plane. We denote $A_0(x)$ the 2-D complex amplitude obtained by the CLOWFS for a perfect wavefront $W(u) = 0$. The complex amplitude $A(x)$ obtained in the CLOWFS detector plane is a linear function of the pupil plane complex amplitude (virtually all optical systems are linear in complex amplitude). Since by definition, $A(x) = A_0(x)$ for $W(u)=0$, we can therefore write 
\begin{equation}
A(x) = A_0(x) + M W(u).
\end{equation}
where $M$ is a linear operator.
The corresponding light intensity $I(x) = |A(x)|^2$ in the CLOWFS detector plane is
\begin{equation}
I(x) =  I_0(x) + 2 Re \left[ A_0(x) \overline{M W(u)} \right] + |M W(u)|^2
\end{equation}
which is linear function of $W(u)$ as long as :
\begin{equation}
\label{equ:lincond}
|M W(u)|^2 \ll Re \left[A_0(x) \overline{M W(u)}\right]
\end{equation}
For small wavefront aberrations, $|M W(u)| \ll |A_0(x)|$, which helps to satisfy condition \ref{equ:lincond}. If there is an aberration mode for which $A_0(x)$ and $M W(u)$ are orthogonal ($Re\left[A_0(x) \overline{M W(u)} \right] = 0$ for all values of $x$), condition \ref{equ:lincond} will not be satisfied even for small aberration levels, and the CLOWFS image $I(x)$ will be a quadratic function of the wavefront aberration ($I(x) = I_0(x) + |M W(u)|^2$). 

Although such a situation is "unlikely" because orthogonality would need to occur simultaneously on each pixel of the CLOWFS detector, it does occur for the focus aberration mode if the CLOWFS detector is conjugated to the focal plane mask (no defocus in the CLOWFS re-imaging). In this special case, since the CLOWFS re-images the focal plane reflective annulus without defocus, we can look at the orthogonality between $A_0(x)$ and $M WF(u)$ directly on the focal plane mask. The complex amplitude $A_0(x)$ is the Fourier transform of the pupil plane complex amplitude and is purely real (no imaginary part) over the reflective annulus of the focal plane mask when no wavefront aberration is present. $M W(u)$ is the change in the complex amplitude over the focal plane annulus introduced by a focus aberration in the pupil plane, and is therefore the Fourier transform of 
\begin{equation}
\label{equ:puppha}
(1- e^{i p(r)}) f(r) \approx i p(r) f(r) 
\end{equation}
where $r$ is the radial coordinate in the pupil plane, $f(r)$ is the amplitude profile in the exit pupil of the PIAA coronagraph ($f(r)$ decreases with $r$), and $p(r)$ is the phase in the exit pupil of the PIAA coronagraph ($p(r)$ is the Focus phase term remapped radially by the PIAA optics). The approximation in Equation \ref{equ:puppha} is valid because we are considering small aberrations ($p(r)<<1$). Since both $p(r)$ and $f(r)$ are real functions, the Fourier transform of $i p(r) f(r)$ is purely imaginary. We have therefore demonstrated that in this special configuration, the aberration-free CLOWFS complex amplitude distribution in the CLOWFS detector array and the change introduced on this complex amplitude by a focus aberration are perfectly orthogonal. This problem can be solved by introducing a defocus in the CLOWFS re-imaging optics, as proposed in this paper. This defocus term is equivalent to convolving the complex amplitude map in the annulus by a kernel which breaks the orthogonality. 

\begin{figure}[htb]
\includegraphics[scale=0.6]{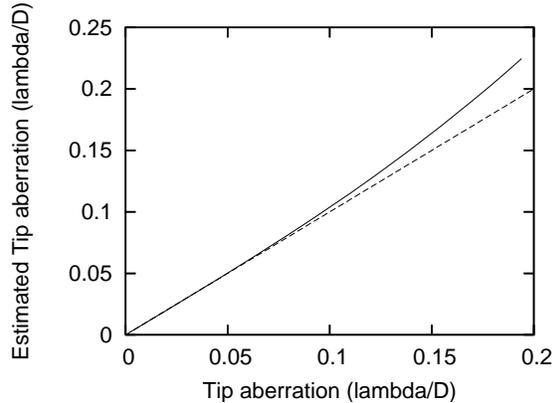} 
\caption{\label{fig:lintip} Linearity of the CLOWFS response to a tilt aberration. With a linear wavefront reconstruction algorithm, the measured Tip signal (solid line) differs slightly from the true Tip aberration (dashed line). This non-linearity reaches about 20\% at 0.2 $\lambda$/D.}
\end{figure}

Numerical simulations show that the CLOWFS's response is linear to the modes considered in this paper. This linearity is only valid for small amplitudes, as shown in Figure \ref{fig:lintip} for tilt. 

\subsection{Wavefront control algorithm and implementation}
\label{ssec:wfca}

\begin{figure*}[htb]
\includegraphics[scale=0.25]{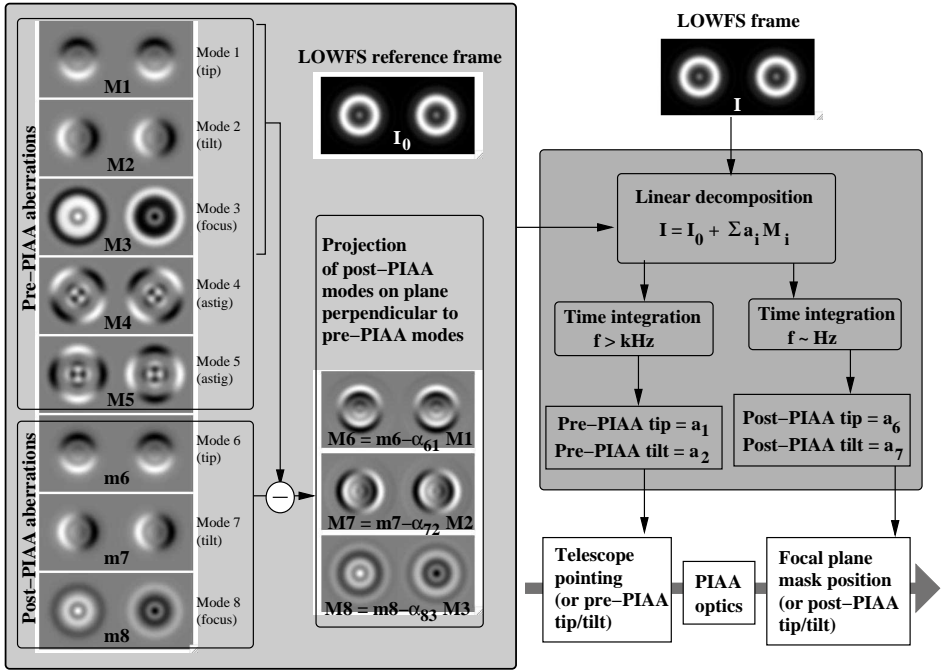} 
\caption{\label{fig:lowfs8mif} Left: CLOWFS calibration prior to sensing. The CLOWFS response to 8 wavefront aberration modes (5 aberrations before the PIAA optics + 3 aberration after the PIAA optics) is measured or computed. The strong similarity between pre-PIAA and post-PIAA modes requires computing "differential aberration" responses M6, M7 and M8. The resulting basis of responses (M1 to M8) is orthogonal, and can then be used to decompose the CLOWFS image in a linear sum of aberrations (top right). Right: An example of a control scheme is shown for tip/tilt modes only (center/bottom right). In this Figure, both inside and outside focus CLOWFS images are shown. The response to Zernike aberrations for inside and outside focus are either equal (for antisymetric Zerike modes) or opposite (for symetric Zernike modes). The CLOWFS sensitivity is therefore identical whether the image acquired is inside or outside focus, or if both images are acquired.}
\end{figure*}

As shown in Figure \ref{fig:lowfs8mif}, the main steps of the CLOWFS closed loop are :
\begin{itemize}
\item{Acquire a single frame with the CLOWFS detector array}
\item{Compute the difference between this frame and the "reference" image obtained (or computed) with no aberrations.}
\item{Decompose this difference as a linear sum of modal responses. These modal responses are either pre-computed or measured prior to starting the CLOWFS loop.}
\item{The coefficients of the linear decomposition described above are used to drive actuators to remove the aberrations measured by the CLOWFS}
\end{itemize}

The estimation of low-order aberrations therefore requires knowledge of the modal responses (how aberrations will linearly modify the image acquired by the CLOWFS). As shown in Figure \ref{fig:lowfs8mif} (left), this information can consist of CLOWFS responses to a series of wavefront aberration modes, which are here Zernike aberrations introduced either before (M1 to M5) or after (m6 to m8) the PIAA optics. They can either be pre-computed by simulation or measured as a response to an aberration introduced in the optical train (this latter method is more robust as it will accurately account for unknown fabrication errors in CLOWFS optical components).

The wavefront control algorithm must avoid confusion which could arise from the fact that the same low order aberration will produce nearly the same CLOWFS signal (although with a different amplitude) wether it is introduced before or after the PIAA optics: in the configuration adopted in this paper, responses to pre-PIAA and post-PIAA tip/tilt are 99.6\% similar. Although post-PIAA aberrations are expected to be smaller/slower, they still need to be properly corrected. One must avoid compensating for a telescope pointing error (pre-PIAA tip-tilt)  by introducing a post-PIAA tip-tilt, or compensating for a post-PIAA tip-tilt aberration by a pre-PIAA tip-tilt correction: both scenarios would create strong diffracted light at and beyond the IWA in the science focal plane even though the "overall" tip-tilt signal seen by the CLOWFS would be zero. 
The reconstruction algorithm shown in Figure \ref{fig:lowfs8mif} addresses this issue by creating "differential" tip-tilt and focus modes (M6, M7, M8) obtained by subtraction of pre-PIAA responses M1,M2, M3 from post-PIAA responses m6, m7, m8. These differential modes track the difference between pre and post-PIAA aberrations, and can only be measured slowly since the corresponding CLOWFS response is weak (due to the high degree of similarity between modes M1, M2, M3 and m6, m7, m8).
Since most tip-tilt/focus aberrations originate before the PIAA optics, the proposed algorithm offloads all of the tip/tilt signal to telescope pointing corrections (pre-PIAA). Differential aberrations are measured more slowly and off-loaded as post-PIAA corrections. In the example given here, thanks the large gap between pre-PIAA and post-PIAA pointing control bandwidths ($\approx$ kHz vs. $\approx$ Hz) the differential pointing signal can be sent to a post-PIAA tip/tilt corrector rather than a combination of pre-PIAA and post-PIAA tip-tilt.


\begin{figure*}[p]
\includegraphics[scale=0.25]{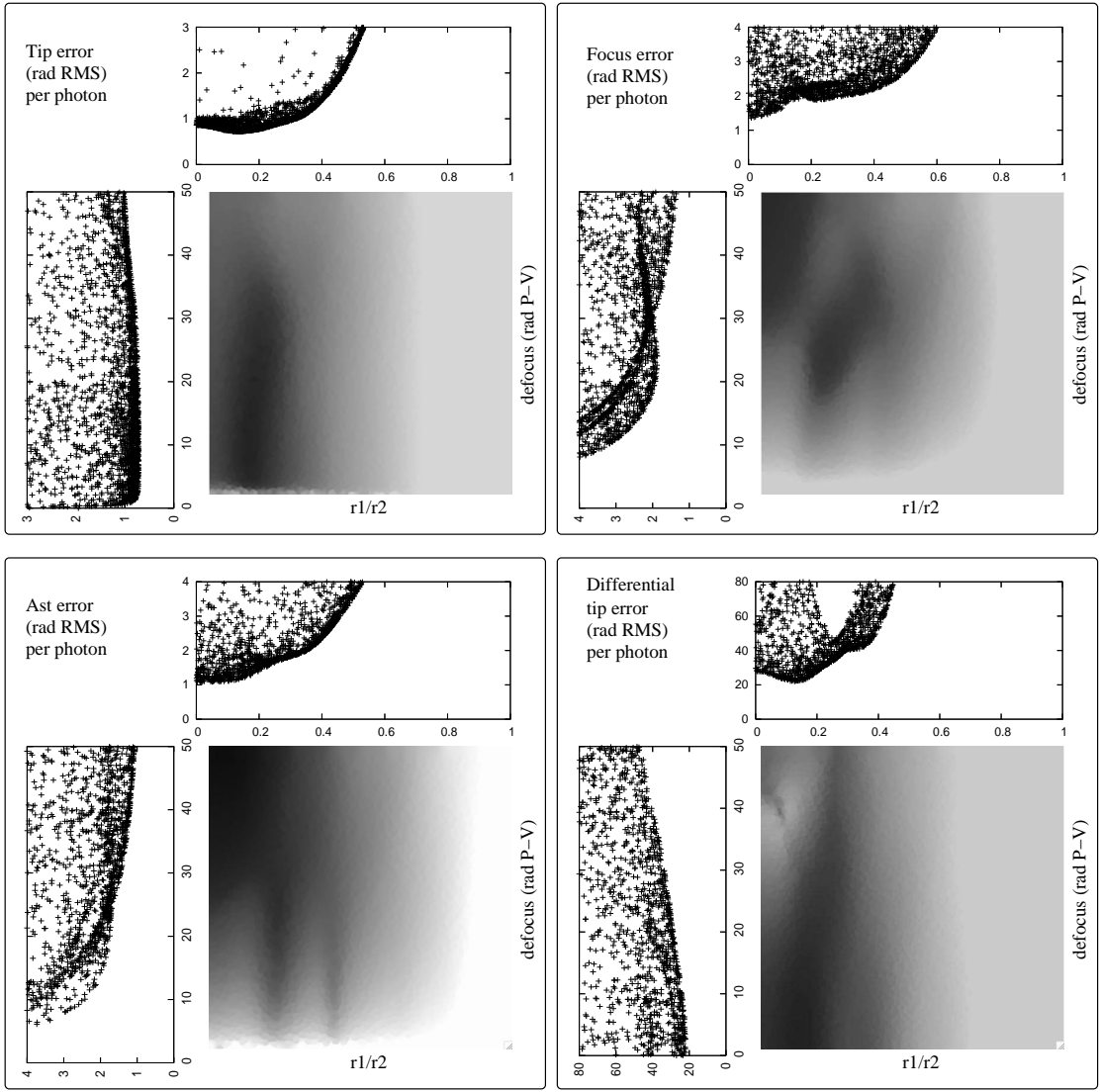} 
\caption{\label{fig:map3d} Tip (top left), focus (top right), astigmatism (bottom left) and differential tip (bottom right) sensitivity of the CLOWFS as a function of the relative size of the opaque disk in the focal plane mask ($r_1/r_2$) and the CLOWFS defocus distance. The sensitivity map is shown as a grey scale 2D map and the corresponding projection on the $r_1/r_2$ and $defocus$ axes are shown as plots above and to the left of each 2D map. Sensitivities are measured as the dispersion on a sample of $10^5$ uncorrelated measurements with $10^6$ photons at the telescope entrance each, and are shown here scaled to one photon (equal to the dispersion multiplied by the square root of the number of photon).}
\end{figure*}

\subsection{Closed loop accuracy and stability}
The accuracy of the CLOWFS loop is ultimately limited by the measurement accuracy (assuming ideal actuators). For small wavefront excursions, the linear model is a very good approximation and the measurement accuracy is driven by detector noise and photon noise. For optimal performance, the CLOWFS exposure time and loop gain need to be carefully chosen to match the dynamic properties of the wavefront aberrations. In principle, in a very stable environment, the CLOWFS loop could correct very small errors by averaging a large number of measurements (small loop gain). Quantitative estimates will be given in \S\ref{sec:aberrsens} assuming a photon-noise limited detector.

The closed loop controller proposed in this paper is based on the linearity of the CLOWFS response. For large wavefront excursions, the actual CLOWFS response can differ significantly from this linear model, and the closed loop may "unlock". While we have not performed a detailed loop stability analysis, we expect the CLOWFS loop to stay locked as long as pointing excursions are within a few tenths of $\lambda/D$:
\begin{itemize}
\item{Figure \ref{fig:lintip} shows that the CLOWFS is still very close to the linear model at 0.2 $\lambda/D$, and, more importantly, the curve shown in this figure is smooth and monotonic. With a 0.2 $\lambda/D$ pointing excursion, the estimate obtained from our linear model will not be exact, but will be sufficiently good to bring back the pointing significantly closer to on-axis at the next iteration}
\item{As illustrated in Figure \ref{fig:lowfsims}, the CLOWFS recovers pointing errors by essentially looking for a brightness enhancement on one side of the re-imaged reflective ring and a corresponding decrease in surface brightness on the opposite side of the ring. As long as pointing errors are small enough for the peak of the stellar image to stay within the inner edge of the reflective ring, this scheme will correctly estimate the direction of the pointing error, and the amplitude of this error will be estimated with sufficient precision to reduce the pointing error after correction. Since the inner edge of the reflective mask is at least 0.5 $\lambda/D$, the pointing loop is expected to stay stable over this range.}
\end{itemize}
Laboratory operation of the CLOWFS confirm this behavior, and the closed loop locks even if the initial pointing error is ~0.5 $\lambda/D$.

If the initial pointing error is too large for the linear CLOWFS loop to lock, a non-linear model of the CLOWFS may be used. Even with a non-linear model, the CLOWFS will fail to measure pointing errors if the stellar PSF misses the focal plane mask (pointing error is as large or larger than the mask size). A separate coarse pointing sensor is then necessary to measure the pointing offset and then bring the stellar PSF within the CLOWFS range. The coronagraph image may be used for this purpose, provided that detector saturation does not prevent pointing error measurement.

\subsection{Optimal CLOWFS design for PIAA coronagraph}
In this section, we explore how the CLOWFS performance is affected by the choice of $r_1/r_2$ (which sets the amount of light sent to the CLOWFS) and the amount of $defocus$ introduced in the image of the focal plane mask. For each choice of the parameters $r_1/r_2$ and $defocus$, the linear algorithm described in \S\ref{ssec:wfca} is used here to measure the CLOWFS sensitivity to pointing (modes M1, M2), focus (mode M3), astigmatism (modes M4, M5), differential tip-tilt (modes M6,M7) and differential focus (mode M8). The linear algorithm allows rapid evaluation of a large number of designs: 2-D sensitivity maps on a $r_1/r_2$ - $defocus$ plane can therefore be created, as shown in Figure \ref{fig:map3d}.

Under the ideal conditions used in this simulation (perfect detector), Figure \ref{fig:map3d} shows that all aberrations are best measured if $r_1/r_2$ is close to zero: ideally, the focal plane mask should be fully reflective. Figure \ref{fig:map3d} however shows that there is relatively little loss in sensitivity if $r_1/r_2$ is increased to up to $\approx 0.4$. With only 10\% of the starlight reflected to the CLOWFS ($r_1/r_2 = 0.294$) there is almost no loss in CLOWFS sensitivity to pointing errors. For the pointing error to double,  $r_1/r_2$ must be increased to 0.44, leaving only 0.87\% of the starlight for the CLOWFS. If the same 0.87\% were applied uniformly over the PSF (instead of masking the center of the PSF), pointing errors would increase more than tenfold.  A similar behavior is observed for all other modes shown in Figure \ref{fig:map3d}: for example, the focus measurement error is doubled for $r_1/r_2 = 0.56$, leaving only 0.06\% of the starlight for the CLOWFS. These results confirm the "relative signal amplification" effect previously claimed: the central dark part of the mask removes most of the CLOWFS flux but only mildly reduces the signal. The CLOWFS design adopted in this paper ($r_1/r_2 = 0.4$, $defocus$ = 20 rad PV) uses only 1.8\% of starlight, at the expense of almost doubling tip-tilt and focus measurement errors.

The amount of $defocus$ has little effect on the CLOWFS sensitivity to pointing, unless it is sufficiently large ($> 40$ rad) to "blend" together light from opposite sides of the reflective focal plane ring. Some $defocus$ is however essential to allow good focus estimation with the CLOWFS, and helps with astigmatism as well. For focus, there is however no gain beyond $defocus \approx 20$ rad, at which point the loss in pointing sensitivity is still very small. 

For all pre-PIAA low order aberrations, Figure \ref{fig:map3d} shows that the residual sensing errors with a CLOWFS are within a factor $\approx$2 of the theoretically optimal sensitivity. For example, the CLOWFS pointing error is $\approx$ 1 rad per mode per photon, or $\approx$ 1.4 rad for tip and tilt combined. This corresponds to just under 1 $\lambda$/D pointing error for a single photon, even if the central opaque zone of the focal plane mask removes most of the starlight. Similarly, Focus and astigmatisms are recovered with $\approx$ 2 rad RMS error each  for a single photon. The CLOWFS therefore makes a very efficient use of a limited number of photons. Due to the strong similarity between pre-PIAA and post-PIAA modes, the CLOWFS sensitivity to differential modes such as M6, M7 or M8, is much weaker.

Beyond the global trends outlined above, Figure \ref{fig:map3d} also shows more subtle effects: diffraction effects produce oscillations of the sensitivity to Focus and Astigmatism in the $r_1/r_2$ - $defocus$ plane. Computing 2D sensitivity maps such as the ones shown in Figure \ref{fig:map3d} is therefore essential for fine tuning the CLOWFS performance to low order wavefront control requirements.

\section{Aberration sensitivity of a PIAA coronagraph equipped with a CLOWFS}
\label{sec:aberrsens}

\subsection{Comparison between CLOWFS and focal plane based wavefront control}

\begin{deluxetable}{l c c}
\tabletypesize{\small}
\tablecaption{\label{tab:perfsumm} Pointing stability requirements for a PIAA coronagraph with and without CLOWFS \tablenotemark{a}}
\tablehead{ & \colhead{Without CLOWFS} & \colhead{With CLOWFS}}
\startdata						
Required pointing calibration accuracy ($10^{-11}$ contrast) &  \multicolumn{2}{c}{0.0016 $\lambda$/D (0.13 mas)}\\
Maximum RMS pointing excursion ($10^{-10}$ contrast) & \multicolumn{2}{c}{0.005 $\lambda$/D (0.4 mas)}  \\
Required sampling time\tablenotemark{b} & 5 s\tablenotemark{c} & 38 $\mu$s \\
Maximum allowed uncalibrated pointing drift rate & 0.026 $mas.s^{-1}$ &  3.4 arcsec/s\\
\enddata
\tablenotetext{a}{For detection of a $10^{-10}$ contrast source around a $m_V = 6$ star observed with a 1.4-m telescope in a 0.2$\mu$m wide band centered at 0.55 $\mu$m with a 50\% system throughput.}
\tablenotetext{b}{Sampling time required to measure the pointing error with a 1-$\sigma$ error equal to the "Required pointing calibration accuracy".}
\tablenotetext{c}{Assumes that 50\% of the observing time is dedicated to measurement of low order aberrations. Also assumes that the signal is well above readout noise and zodi/exozodi background.}
\end{deluxetable}

We consider here a 1.4-m diameter telescope observing a $m_V = 6$ star in a 0.2 $\mu$m wide band centered at $\lambda = 0.55 \mu m$. We assume a 50\% system throughput, which corresponds to $6\times10^{9} \: ph.s^{-1}$ at the telescope entrance. As described in \S\ref{sec:requ}, pre-PIAA low order aberrations can affect coronagraphic contrast at small angles (close to the IWA of the PIAA coronagraph) much more easily than post-PIAA low order aberrations. We therefore only consider in this section pre-PIAA pointing errors, which could be generated by a telescope/spacecraft pointing error. We adopt the requirement defined in \S\ref{sec:requ}: pointing error must be measured to a 0.0016 $\lambda/D$ 1-$\sigma$ accuracy.

Without CLOWFS, a 0.0016 $\lambda$/D pointing error would have to be measured from the corresponding $6.9\times10^{-11}$ total coronagraphic leak in the science focal plane, equal to approximately $0.4\: ph.s^{-1}$. A 1-$\sigma$ measurement of this leak can be achieved with $\approx$ 1 photon, provided that this leak is interferometrically combined with a much brighter and well known "reference(s)" (in 2-D, this technique is referred to a Focal plane wavefront sensing, where a Deformable Mirror is used to mix coherent starlight with the scattered light halo). This "reference(s)" is necessary to measure the sign of the aberration and also to bring the signal above the detector readout noise and/or incoherent background in the image. In this scheme, detection of a faint companion cannot be done at the same time as pointing measurement, and time must be shared between the two tasks (we assume here that 50\% of the time is spent for each task). A measurement of pointing error with a 0.0016 $\lambda/D$ 1-$\sigma$ error would therefore require 5 seconds.

As shown in Figure \ref{fig:map3d}, the tip sensitivity for the CLOWFS is 1.2 rad RMS for one photon at the telescope entrance. The 0.0016 $\lambda/D$ tip measurement accuracy (equal to 0.0025 rad RMS) therefore requires $2.3 \times 10^5$ photons, which can be gathered in 38 $\mu$s. The CLOWFS is therefore capable of measuring pointing errors about 130,000 times faster than the science focal plane. 

The pointing stability requirement is derived in table \ref{tab:perfsumm} from both the required pointing calibration accuracy (equal to 0.0016 $\lambda/D$ in this example) and the sampling time necessary to measure pointing errors to this level of accuracy. The ratio between these two quantities defines a maximum allowable pointing drift rate beyond which the stability/calibration requirement previously defined cannot be met.
As shown in Table \ref{tab:perfsumm}, measuring pointing errors with the CLOWFS is several orders of magnitude quicker than if only the light in the science focal plane were used. This measurement sensitivity gain yields a much more relaxed pointing drift rate requirement: $3.4 arcsec.s^{-1}$ with a CLOWFS as opposed to $0.026 \:mas.s^{-1}$ without.

\subsection{Detector requirements}
The CLOWFS camera requires a modest number of pixels: the signals shown in the left of Figure \ref{fig:lowfs8mif} contain little high spatial frequencies and can be accurately measured with approximately 10 pixels across one defocused spot images. A 20 by 20 pixel window (400 pixels) is sufficient, and can be read rapidly ($>$10 kHz) with current technology.

The CLOWFS measures pointing and other low order aberration by detecting changes in the defocused image it acquires. Temporal changes in the detector response must therefore be small compared to the expected signal. The CLOWFS is not sensitive to static spatial variations in the detector response (flat field), as the signal is extracted from a difference between two images acquired at different times. The effect of spatially uniform flat field variations can be removed by scaling of the images prior to this subtraction.
A 0.0016 $\lambda$/D pointing offset corresponds to a 2\% change in the surface brightness on the bright ring images by each frame of the CLOWFS: one side of the ring is 2\% brighter while the opposite side is 2\% fainter. A comparable variation in detector sensitivity between the two sides of the ring images is very unlikely in modern detectors (visible CCDs), even over the course of several hours.

Thanks to the large number of photons collected by the CLOWFS, a moderate amount of readout noise (few photo-electron) will still allow operation of the CLOWFS at high sampling rate.  For example, at 10 kHz sampling rate, the defocused image contains 10800 photons on a $m_V = 6$ target. Assuming that the "ring" occupies a 50 pixel surface area on the detector, photon noise (15 photo-electrons per pixel) is likely to be larger than readout noise on modern visible detectors.

\section{Laboratory Demonstration}
\label{sec:labexp}

\begin{figure}[htb]
\includegraphics[scale=0.3]{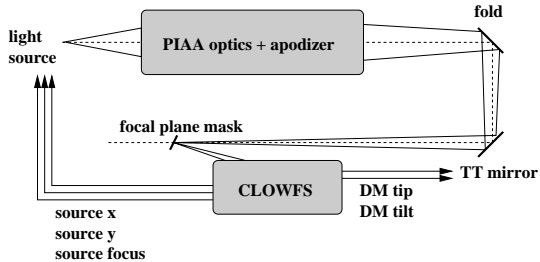} 
\caption{\label{fig:labsetup}Schematic representation of the CLOWFS implementation for the PIAA coronagraph testbed at Subaru Telescope. The CLOWFS is extracting light reflected by the focal plane mask and drives 5 actuators: pre-PIAA tip/tilt/focus (the light source is mounted on a 3 axis piezo stage) and post-PIAA tip/tilt.}
\end{figure}

The CLOWFS has been implemented in the PIAA coronagraph testbed at the Subaru Telescope. As shown in Fig. \ref{fig:labsetup}, the CLOWFS is driving 5 piezo actuators to move the light source in x,y,z and move a post-PIAA mirror in tip and tilt. 

\begin{figure}[htb]
\includegraphics[scale=0.3]{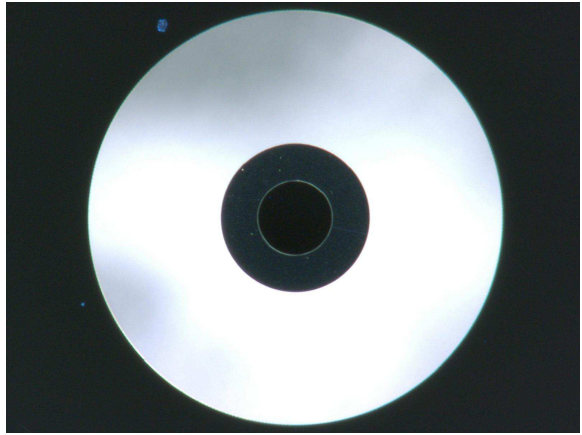} 
\caption{\label{fig:labmask} CLOWFS focal plane mask used in the PIAA coronagraph laboratory testbed at Subaru Telescope (fabricated by HTA photomask). The 100 micron radius mask center is opaque (low reflectivity), and is surrounded by a 100 micron wide highly reflective annulus. The science field, transmiting light to the science camera, extends from 200 micron to 550 micron radius.}
\end{figure}

\begin{figure*}[htb]
\includegraphics[scale=0.28]{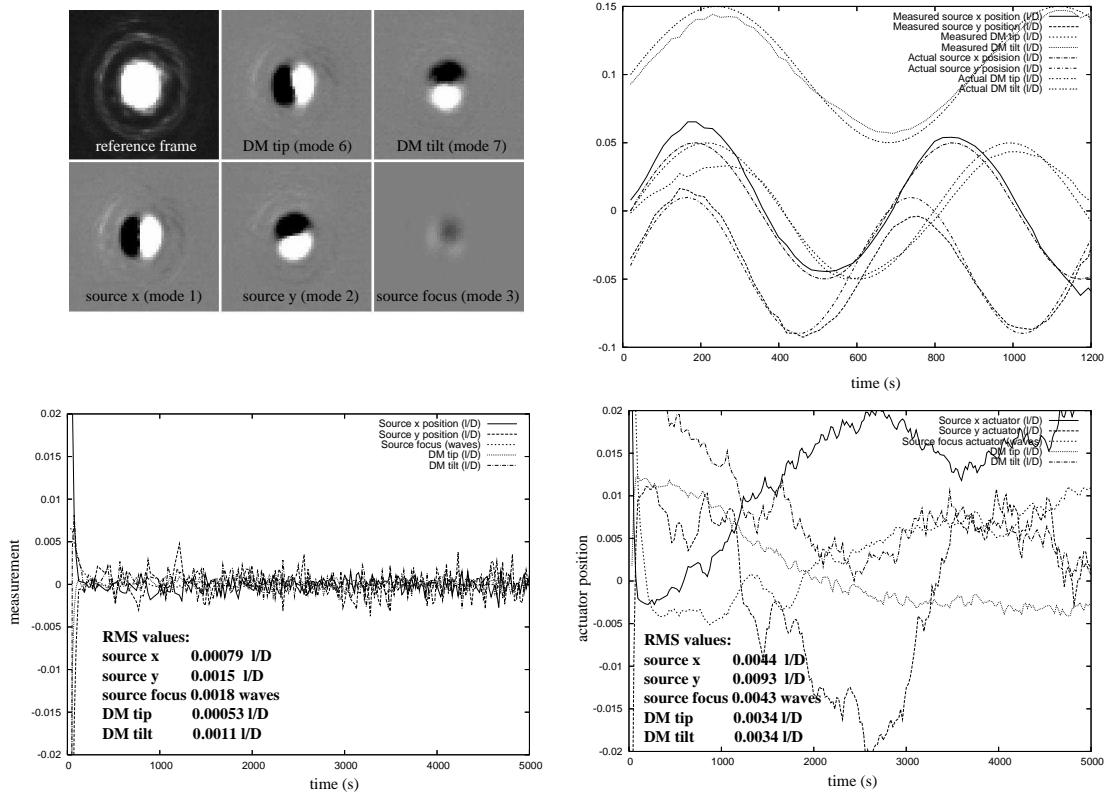} 
\caption{\label{fig:labplots} Laboratory performance for the CLOWFS. Upper left: Measured CLOWFS reference frame and influence functions for the 5 axis controlled in the experiment. Pre-PIAA and post-PIAA modes look extremely similar, as expected. Top right: Open loop simultaneous measurement of pre and post-PIAA modes. The measured amplitudes match very well the sine-wave signals sent to the actuators, and the CLOWFS is able to accurately measure all 4 modes shown here with little cross-talk. Since this measurement was performed in open loop, the measurement also include unknown drifts due to the limited stability of the testbed. Bottom left: Closed loop measurement of the residual error for the 5 modes controlled. The achieved pointing stability is about $10^{-3}$ $\lambda/D$ for both the pre-PIAA and post-PIAA tip/tilt. Bottom right: Position of the actuators during the same closed loop test.}
\end{figure*}

The CLOWFS focal plane mask used in this experiment is shown in Fig. \ref{fig:labmask}, and was manufactured by lithography techniques. The mask is not ``ideal'': the central portion is not perfectly opaque, and reflects a few percent of the light. Since most of the starlight falls on this central part of the mask, the CLOWFS frame contains a bright peak at the center of the defocused mask image. Thanks to the defocus introduced in the CLOWFS, the fainter light reflected by the reflective annulus interferes with this central peak, and also forms fainter outer rings visible as shown in Fig. \ref{fig:labplots}. With a finite reflectivity of the central part of the focal plane mask, a more optimal design would be to reduce the size of the central ``dark'' area to increase the amount of light at the transition between the central dark spot and the reflective annulus. 

The linear scheme implemented for calibration and processing of the CLOWFS data is very insensitive to the details of the CLOWFS design, and the CLOWFS is able to measure simultaneously both pre and post-PIAA tip/tilt with little cross-talk (see Fig. \ref{fig:labplots}, upper right). As shown in Fig. \ref{fig:labplots}, we have achieved $\approx 10^{-3} \lambda/D$ closed loop pointing stability for both the light source position and post-PIAA tip/tilt.

In our laboratory testbed, pre-PIAA pointing errors and post-PIAA tip/tilt are similar in amplitude, while in an actual system, we would expect pre-PIAA pointing errors (telescope pointing, primary and secondary mirrors tilts) to be much larger and faster than post-PIAA tip/tilt (most likely due to slow thermal drifts on small optics). On our testbed, we have therefore operated the control loop with similar temporal bandwidth for both pre and post-PIAA aberrations, unlike the control scheme proposed in Fig. \ref{fig:lowfs8mif}.

The performance we have achieved in the laboratory is limited by both system stability (our testbed is in air and includes 75 mm diameter optics separated by more than a meter) and CLOWFS loop speed. In our experiment, the CLOWFS sampling interval was limited to 25 s due to hardware limitations (readout speed of the camera image and time necessary to transfer the image from the computer which controls the camera to a separate computer performing the CLOWFS image analysis).
For a space coronagraphic mission, a better controlled environment and a faster readout camera (10 kHz is reasonable for the small number of pixels needed) would allow higher performance. Despite these limitations, our laboratory demonstration of the CLOWFS concept has exceeded both the RMS pointing excursion and the pointing measurement accuracy required for achieving 1e10 contrast at 2 $\lambda/D$ with a visible PIAA coronagraph mission.

\section{Conclusion}
The CLOWFS design presented in this paper can efficiently measure low order aberrations "for free", as it uses light that would otherwise be discarded by the coronagraph. Both the hardware configuration and software algorithms presented are easy to implement and their performance is robust against calibration errors, chromaticity, non-common path errors and small errors/aberrations in the optical components. The CLOWFS pointing measurement can also lead to improved astrometric accuracy for the position of faint companions, as the star position on the image is usually difficult to measure in coronagraphic images \citep{digb06}.

In this paper, we have studied a CLOWFS design on a low-IWA PIAA coronagraph. Although coronagraphs with larger IWA can tolerate larger amount of low order aberrations (see for example Kuchner \& Traub 2002; Kuchner et al. 2005), they require these aberration to be measured ahead of the coronagraphed beam because they are "blind" to aberrations until they are large enough to produce significant coronagraph leaks. The CLOWFS would therefore be very useful to any high contrast coronagraph, and the optical design presented in this paper can readily be used on any coronagraph where a focal plane mask physically blocks starlight.

The CLOWFS can also be applied to phase mask coronagraphs. For such coronagraphs (see for example Roddier \& Roddier 1997; Rouan et al. 2000; Palacios 2005), where starlight is diffracted outside the pupil by the focal plane mask, a modified Lyot stop is placed on the pupil plane to reflect starlight to the CLOWFS. In addition, using a pattern matching algorithm, the CLOWFS can estimate low-order wavefront aberrations accurately and quickly even in non-linear region. These new CLOWFS designs and their performances will be presented in an upcoming paper.

\end{document}